\documentclass[%
reprint,
superscriptaddress,
 amsmath,amssymb,
 aps,
]{revtex4-1}

\usepackage{graphicx}
\usepackage{dcolumn}
\usepackage{bm}

\usepackage{todonotes}
\usepackage{xcolor}
\usepackage[normalem]{ulem}
\usepackage{braket}
\usepackage{tikz}
\usepackage{changes}

\newcommand{\rev}[1]{\textcolor{black}{#1}}

\newcommand{\SP}{single-particle}
\newcommand{\SO}{spin-orbit}

\usepackage[pagebackref=false]{hyperref}

\definecolor{myblue}{RGB}{0, 112, 192}

\DeclareUnicodeCharacter{2212}{-}

\begin{document}

\author{S.~M\"oller}
\thanks{These two authors contributed equally.}
\author{L.~Banszerus}
\thanks{These two authors contributed equally.}
\affiliation{JARA-FIT and 2nd Institute of Physics, RWTH Aachen University, 52074 Aachen, Germany,~EU}%
\affiliation{Peter Gr\"unberg Institute  (PGI-9), Forschungszentrum J\"ulich, 52425 J\"ulich,~Germany,~EU}
\author{A. Knothe}
\affiliation{National Graphene Institute, University of Manchester, Manchester M13 9PL, United Kingdom}%
\author{C.~Steiner}
\affiliation{JARA-FIT and 2nd Institute of Physics, RWTH Aachen University, 52074 Aachen, Germany,~EU}
\author{E.~Icking}
\affiliation{JARA-FIT and 2nd Institute of Physics, RWTH Aachen University, 52074 Aachen, Germany,~EU}%
\affiliation{Peter Gr\"unberg Institute  (PGI-9), Forschungszentrum J\"ulich, 52425 J\"ulich,~Germany,~EU}


\author{S.~Trellenkamp}
\author{F.~Lentz}
\affiliation{Helmholtz Nano Facility, Forschungszentrum J\"ulich, 52425 J\"ulich,~Germany,~EU}

\author{K.~Watanabe}
\affiliation{Research Center for Functional Materials, 
National Institute for Materials Science, 1-1 Namiki, Tsukuba 305-0044, Japan}
\author{T.~Taniguchi}
\affiliation{International Center for Materials Nanoarchitectonics, 
National Institute for Materials Science,  1-1 Namiki, Tsukuba 305-0044, Japan}%
\author{L.~I.~Glazman}
\affiliation{Departments of Physics and Applied Physics, Yale University, New Haven, CT 06520, USA}
\author{V.~I.~Fal'ko}
\affiliation{National Graphene Institute, University of Manchester, Manchester M13 9PL, United Kingdom}%
\affiliation{Department of Physics and Astronomy, University of Manchester, Oxford Road, Manchester, M13 9PL, United Kingdom}
\affiliation{Henry Royce Institute for Advanced Materials, University of Manchester, Manchester, M13 9PL, United Kingdom}
\author{C.~Volk}
\affiliation{JARA-FIT and 2nd Institute of Physics, RWTH Aachen University, 52074 Aachen, Germany,~EU}%
\affiliation{Peter Gr\"unberg Institute  (PGI-9), Forschungszentrum J\"ulich, 52425 J\"ulich,~Germany,~EU}
\author{C.~Stampfer}
\email{stampfer@physik.rwth-aachen.de}
\affiliation{JARA-FIT and 2nd Institute of Physics, RWTH Aachen University, 52074 Aachen, Germany,~EU}%
\affiliation{Peter Gr\"unberg Institute  (PGI-9), Forschungszentrum J\"ulich, 52425 J\"ulich,~Germany,~EU}%

\title{Probing two-electron multiplets in bilayer graphene quantum dots}

\date{\today}


\begin{abstract} 
We report on finite bias spectroscopy measurements of the two-electron spectrum in a gate defined bilayer graphene (BLG) quantum dot for varying magnetic fields. The spin and valley degree of freedom in BLG give rise to  multiplets of 6 orbital symmetric and 10 orbital anti-symmetric states. We find that orbital symmetric states are lower in energy and separated by $\approx 0.4 - 0.8$~meV from orbital anti-symmetric states. The symmetric multiplet exhibits an additional energy splitting of its 6 states of $\approx 0.15 - 0.5$~meV due to lattice scale interactions. The experimental observations are supported by theoretical calculations, which allow to determine that inter-valley scattering and 'current-current' interaction constants are of the same magnitude in BLG.


\end{abstract}

\maketitle

 \begin{figure}[!htb]
\includegraphics[draft=false,keepaspectratio=true,clip,width=\linewidth]{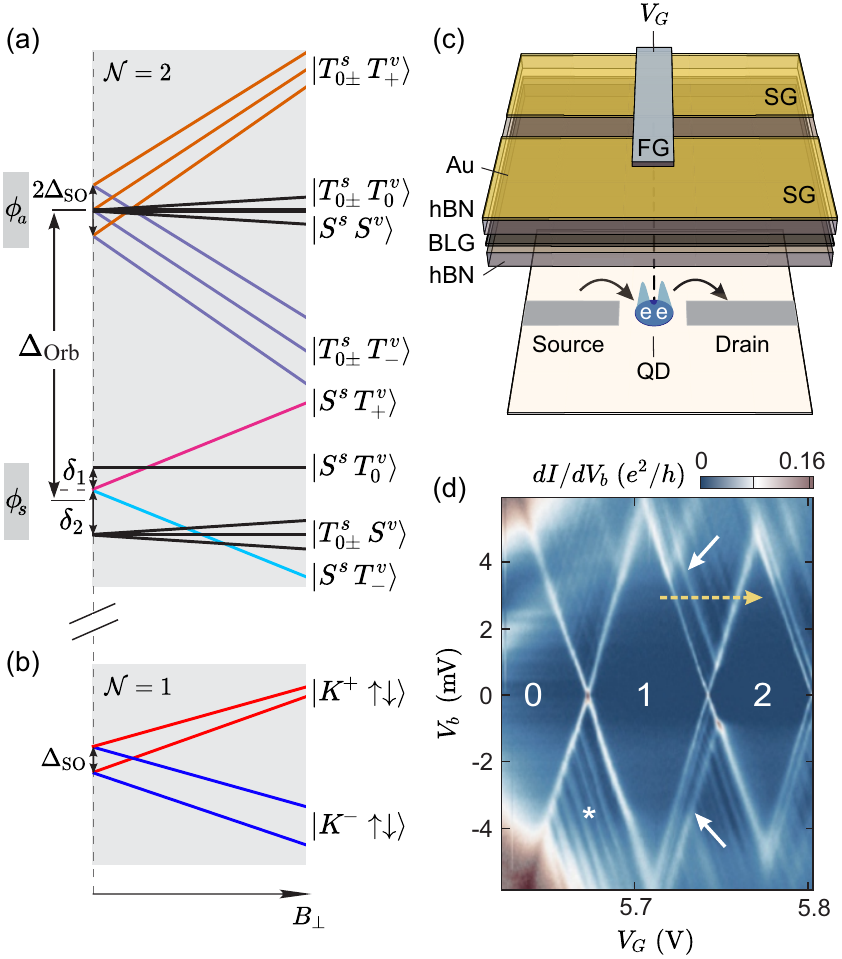}
\caption[Fig01]{
\textbf{(a, b)} Single-, $\mathcal{N} = 1$, and two-particle, $\mathcal{N} = 2$, energy spectrum of a BLG QD in \rev{an out-of-plane} 
magnetic field. The two-particle states are split into two multiplets with (anti-) symmetric orbital wavefunction, ($\phi_a$) $\phi_s$, consisting of (10) 6 states. They are separated by an orbital splitting, $\Delta_\mathrm{Orb}$, while the symmetric multiplet is split again, denoted with $\delta_{1,2}$. The valley $g$-factor is different for single-particle, symmetric and anti-symmetric orbitals.
\textbf{(c)} Schematic of the device with BLG encapsulated in hexagonal boron nitride and multiple gate layers that create the QD via soft-confinement.
\textbf{(d)} Differential conductance $d\text{I}/d\text{V}_b$ with respect to applied bias and FG voltage, showing diamond shaped regions of Coulomb blockade. Finite bias measurements are performed along the yellow arrow to inspect the excited state spectrum in the single and two-particle regime. \rev{White arrows highlight excited states of the two-particle spectrum.}
}
\label{f1}
\end{figure}

Graphene quantum dots (QDs) are considered promising candidates for spin-based quantum computation, as the low spin-orbit and hyperfine coupling provides long spin coherence times~\cite{Loss1998Jan, Bennett2000Mar, Trauzettel2007Mar, Silvestrov2007Jan, Rohling2012Aug}. In addition to the spin, graphene offers the valley degree of freedom, which gives rise to a rich energy spectrum and creates the opportunity for the implementation of valley and Kramer's qubits~\cite{Rohling2014Oct, Rohling2012Aug, Laird2015Jul}.
Recent experimental progress on electrostatically confined bilayer graphene (BLG) QDs, demonstrating single-electron occupation~\cite{Eich2018Jul, Banszerus2020Mar, Banszerus2020Oct}, gate-tunable valley $g$-factors~\cite{Tong2021Jan} and low spin-orbit coupling~\cite{Banszerus2020May,Banszerus2021Sep,Kurzmann2021Mar}, brings graphene based qubits within reach. 
%
%
As two-electron states are particularly interesting for the implementation of well-controllable qubits, such as exchange and singlet-triplet  qubits~\rev{\cite{Medford2013Jul,Petta2005Sep}}, which offer various advantages over single-electron qubits, a detailed understanding of the two-particle spectrum in BLG QDs is becoming increasingly important. 


%
This is all the more true since the spin and valley ($K^+$, $K^-$) degrees of freedom in BLG yield a total of 16 two-particle states where the wavefunction-dependent valley $g$-factors give rise to a rich level spectrum.
The total two-particle wavefunction in BLG can be factorized into an orbital, a spin and a valley term~\cite{Knothe2020Jun, Knothe2021Apr}, resulting in 6 states with an anti-symmetric spin-valley and a {\it symmetric orbital} wavefunction, and
10 states with a symmetric spin-valley and an {\it anti-symmetric orbital} wavefunction. This gives rise to the symmetric and anti-symmetric multiplet structure of the two-electron spectrum in BLG.

In this letter, we report on the experimental observation and detailed description of the symmetric and anti-symmetric two-electron multiplets in BLG QDs. We confirm the Berry curvature induced wavefunction dependence of the valley magnetic moment, observing different valley $g$-factors for single-particle, symmetric and anti-symmetric orbital wavefunctions, which gives rise to a rich magnetic-field dependent transition structure.
Additionally, we show that the energy splitting between the multiplets ($\Delta_\mathrm{Orb}$) can be tuned in the range of $0.4-0.8$~meV and that the splitting of the orbital-symmetric multiplet allows to quantify inherent lattice scale interaction constants in BLG \cite{lemonik_competing_2012, lemonik_spontaneous_2010, aleiner_spontaneous_2007}.

We reconstruct the two-electron spectrum of the BLG QD from finite bias spectroscopy measurements of the $\mathcal{N} = 1 \rightarrow 2 $ transition. 
Therefore, it is necessary to  consider both single- and two-particle states, which are presented in Figs.~1(a,b). Fig.~1(b) shows the energy of the four single-particle states of the lowest orbital as a function of an out-of-plane magnetic field, $B_\perp$. At zero magnetic field, the four states are split into two Kramers pairs by Kane-Mele spin-orbit coupling, $\Delta_\mathrm{SO}$,~\cite{Kane2005Nov,Banszerus2020May,Banszerus2021Sep,Kurzmann2021Mar}. At finite magnetic field, the states shift linearly in energy according to their spin and valley Zeeman effect $\Delta E(B_\perp) = \frac{1}{2}(\pm g_s \pm g^{(1)}_v) \mu_B B_\perp$, with the spin $g$-factor $g_s=2$, the single-particle (wavefunction dependent) valley $g$-factor $g^{(1)}_v$ and the Bohr magneton $\mu_B$ \cite{Knothe2018Oct, Knothe2020Jun, Eich2018Aug,Banszerus2020Dec}. Note that the valley $g$-factor is usually around one order of magnitude larger than the spin $g$-factor. Measurements confirming the well-understood single-electron spectrum in our single QD are shown in Supplementary Fig.~S2. 
%
%
%
%
%
%
%
%
%
The 16 two-particle states are shown in Fig.~1(a), which, at $B_\perp=0$, group into the orbital symmetric ($\phi_s$) and anti-symmetric ($\phi_a$) multiplets separated by the energy $\Delta_\mathrm{Orb}$. For both the valley and spin wavefunction components, there are three symmetric and one anti-symmetric two-particle states, namely the triplets $\ket{T^{s,v}_{0\pm}}$ and the singlet $\ket{S^{s,v}}$, where $s$ and $v$ refer to spin or valley.
There are 6 (10) anti-symmetric (symmetric) combinations of the spin and valley components, which need to be combined with an (anti-)symmetric orbital component for the total wavefunction to remain anti-symmetric~\cite{Pei2012Oct}.
The orbital energy of the two-particle states comprises the occupied \SP{} orbitals' energies and Coulomb interactions \rev{between the two particles}, which lead to a splitting of symmetric and anti-symmetric orbitals by $\Delta_\mathrm{Orb}$, depending on the size of the QD, surrounding screening and applied displacement field \cite{Knothe2020Jun}.
\rev{Additionally, the orbitally symmetric 
states are affected by short-range Coulomb interactions inherent to BLG \cite{lemonik_competing_2012, lemonik_spontaneous_2010, aleiner_spontaneous_2007}, since their orbital wavefunctions have non-zero density at the relevant short inter-particle distances. BLG exhibits local density fluctuations which are determined by the lattice symmetry. Mutual interactions between these fluctuations induce the formation of states with spontaneously broken symmetries in sublattice and valley space, introducing splittings $\delta_{1,2}$ proportional to the strength of the corresponding short-range interactions~\cite{Knothe2021Apr}.} 
The energy of the spin- and valley-dependent part is determined by the coupling to the magnetic field and the \SO{} coupling. Kane-Mele \SO{} coupling induces opposite spin splittings in the two valleys, which only affects states in the anti-symmetric orbital which are both valley and spin polarized. A perpendicular magnetic field couples to spin- and/or valley polarized states $\ket{T^{v, s}_{0\pm}}$ and shifts their energies according to their corresponding $g$-factors. 
Note, that the two-particle valley $g$-factors ($g_v^s, g_v^a$) are in general also different from the single-particle $g$-factor ($g^{(1)}_v$)~\footnote{The simplest $\phi_s$ would be that both charge carriers are in the lowest energy single-particle orbital. In that case, the valley $g$-factors would be the same for the symmetric two-particle and single-particle wavefunctions. Conversely, a differing $g$-factor indicates that also higher energy single-particle orbital states are included in $\phi_s$.}.

\begin{figure*}[!thb]
\centering
\includegraphics[draft=false,keepaspectratio=true,clip,width=\linewidth]{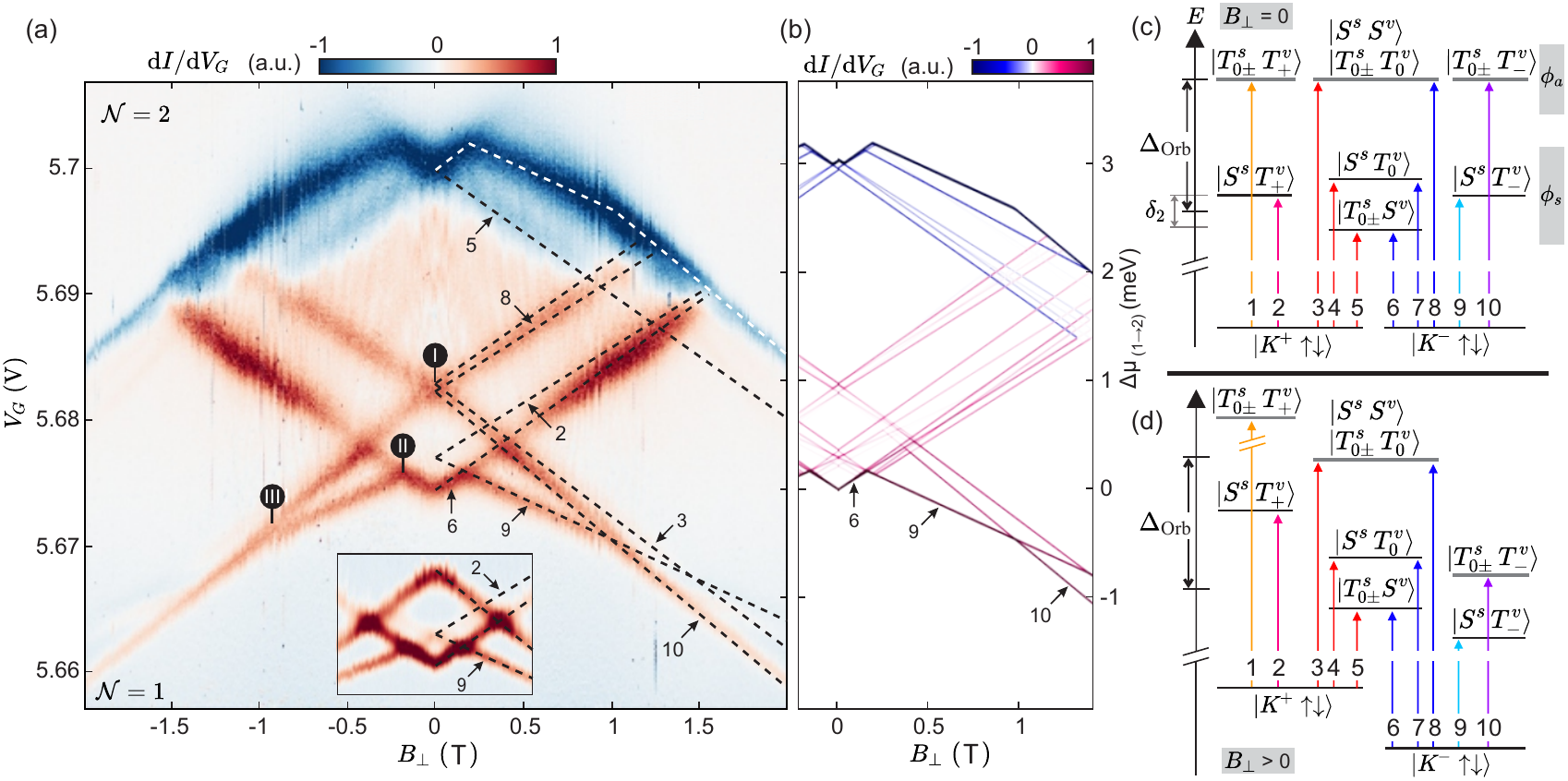}
\caption[Fig02]{
 \textbf{(a)} Differential transconductance $dI/dV_{G}$ as a function of $V_\mathrm{G}$ and $B_\perp$, measured across the $\mathcal{N} = 2$ Coulomb peak at $V_{b} = 3$~mV as indicated in Fig.~1(b). Transitions to states with symmetric and anti-symmetric orbital wavefunction with different $g$-factors give rise to a rich spectrum. The inset shows a zoom-in for low magnetic field with reversed bias, where the multiplet splitting is visible. Prominent features are highlighted by dashed lines, obtained from theoretically predicted features in \textbf{(b)}: Calculated differential conductance map as in (a), assuming $g^{(1)}_v = 39$, $g^s_v = 32$, $g^a_v = 43$, $\Delta_\mathrm{SO} = 80 \, \mu$eV, $\delta_1 = 66\, \mu$eV, $\delta_2 = 306\, \mu$eV and an asymmetry of the tunnel barriers to source $t^S$ and drain $t^D$ of $t^S/t^D = 0.7$. 
 \textbf{(c,d)} Transition scheme from one to two electrons in the QD at zero (c) and finite (d) $B_\perp$, with spin-orbit coupling and spin Zeeman splittings omitted for simplicity. The length of the transition arrows corresponds to the chemical potential necessary to enter the second electron in the QD, which also corresponds to the $y$-axis in panel (a) and (b). 
}
\label{f3}
\end{figure*}

Fig.~1(c) shows a schematic of our QD device, which consists of a BLG flake encapsulated in hexagonal boron nitride (hBN).
This heterostructure is placed on a graphite back gate (BG) and has two  Cr/Au gate layers evaporated on top, a set of split gates (SG) and 
a finger gate (FG). 
To form a QD, we follow previous works~\cite{Eich2018Aug, Eich2018Jul, Banszerus2018Aug, Banszerus2020Mar, Banszerus2020Oct, Banszerus2021Sep, Tong2021Jan, Kurzmann2019Jul}: A narrow p-doped channel is created utilizing SG and BG. 
%
The FG locally overcompensates the potential applied by the BG to form an n-type QD. It  is separated from source and drain by two tunneling barriers, where the Fermi energy resides within the band gap. With a simple plate capacitor model \footnote{We use $E_C = e^2 \cdot d_\mathrm{hBN} / (\pi \epsilon_0 \epsilon_\mathrm{hBN} r^2_\mathrm{QD})$, with the charging energy $E_C$, the dielectric permittivity of hBN, $\epsilon_{hBN} \approx 4$, the thickness of the hBN, $d_\mathrm{hBN}$ and the radius of the QD, $r_\mathrm{QD}$.}, we estimate the radius of our QD to be $\approx 80$~nm, which is compatible with the dimensions of the gate layout.

Fig.~1(d) shows the differential conductance, $dI/dV_\mathrm{b}$, through the QD as a function of the bias, $V_\mathrm{b}$, and FG voltage, $V_\mathrm{G}$, for an electron occupation, $\mathcal{N}$, between zero and two. Within the Coulomb diamonds, $\mathcal{N}$ is fixed and transport is suppressed by Coulomb blockade. The outline of the conducting region between $\mathcal{N} = 1$ and $\mathcal{N} = 2$ is defined by the ground state (GS) transition entering and leaving the bias window. Here, 'GS transition' refers to the QD being in its \rev{GS} before and after the tunneling of the second electron onto the QD. Additional transitions involving excited states (ES) of the $\mathcal{N} = 1$ and/or $\mathcal{N} = 2$ spectrum are possible within that region, as highlighted by the white arrows. Apart from QD transitions, we also observe additional features in the differential conductance in Fig.~1(d) \footnote{All features originating from the QD should be symmetric in bias as long as tunnel barriers are not too asymmetric.}, which most likely originate from density of states effects in the leads~\cite{Manna2017Dec,Thomas2021May} (see white asterisk) and have previously been observed in similar devices~\cite{Eich2018Aug}.


To experimentally investigate the two-particle spectrum, we measure line cuts along the dashed arrow in Fig.~1(d) and inspect the magnetic field dependence of the GS and ES transitions. In Fig.~2(a), we plot the transconductance, $dI/dV_\mathrm{G}$, at $V_\mathrm{b}=3$~mV as function of $V_\mathrm{G}$  and $B_\perp$. The gate voltage along the dashed arrow in Fig.~1(d) was converted to electro-chemical potential (displayed on the right axis of Fig.~2(b)) and the raw data was corrected for magnetic field dependent oscillations in the lever-arm, which are due to Shubnikov-de-Haas oscillations in the lead region~\cite{Banszerus2020Dec}. With increasing $V_G$, first the GS transition enters the bias window, then additional transitions follow, each appearing as feature of increased differential transconductance, giving rise to a rich spectrum. Eventually, the GS transition leaves the bias window, which Coulomb blockades the QD and appears as a feature of strong negative transconductance (white dashed line). The inset shows the reversed bias direction for low magnetic fields \footnote{Due to slight asymmetries in the applied bias, probably due to the tunnel barriers, different features are more prominent at reversed bias.}. \rev{From the linewidth of the features, we may estimate a lower bound for possible spin- and/or valley mixing of $< 100\, \mu$eV, in good agreement with recent measurements on BLG double QDs~\cite{Banszerus2021Sep}.}

\begin{figure*}[!thb]
\centering
\includegraphics[draft=false,keepaspectratio=true,clip,width=0.95\linewidth]{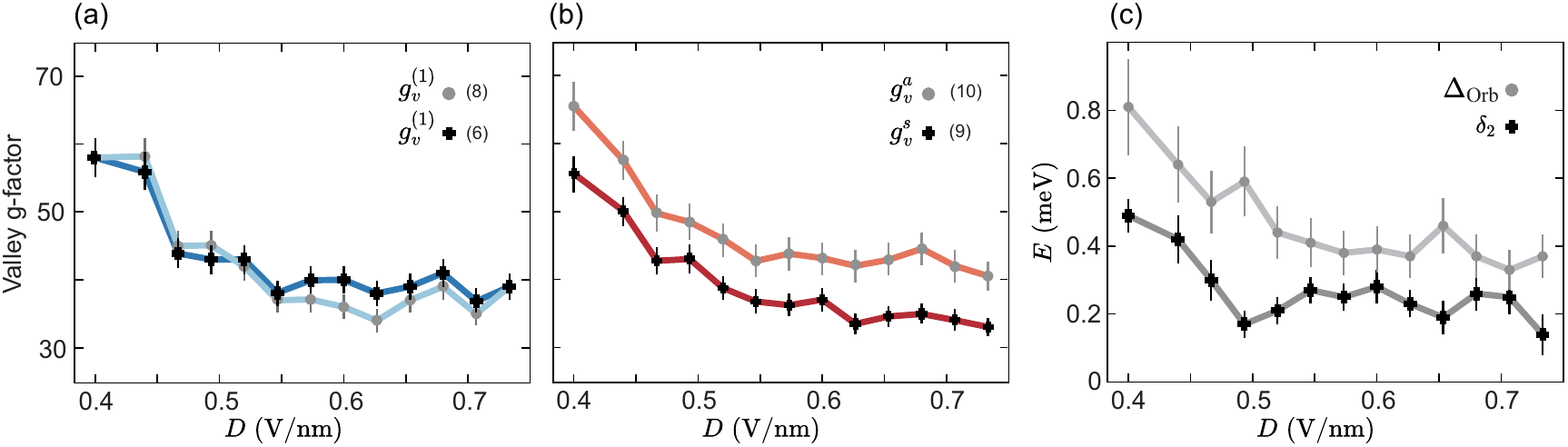}
\caption[Fig03]{
\textbf{(a)} Single-particle $g$-factors, $g^{(1)}_v$, as a function of applied perpendicular displacement field, evaluated from transitions to orbitally symmetric (6) or anti-symmetric (8) two-particle states with vanishing valley magnetic moment (labeling as in Fig.~2(c)). 
\textbf{(b)} Two-particle $g$-factors, $g^{s,a}_v$, evaluated from the transitions to valley triplet states in the (anti-)symmetric orbital. Symmetric and anti-symmetric orbitals exhibit an average $g$-factor difference of $\approx 8$.
\textbf{(c)} Energy splitting between symmetric and anti-symmetric orbital, $\Delta_\mathrm{Orb}$, and multiplet splitting, $\delta_2$, in the symmetrical orbital as a function of displacement field.}
\label{f3}
\end{figure*}

We reproduce the features of Fig.~2(a) in theoretical calculations, which are shown in Fig.~2(b). Tunnelling transport through the QD is described by solving the rate equations for the single- and two-particle states presented in Fig.~1(a) and (b) \footnote{The model for the QD, the leads, and the QD-lead tunnel coupling are presented in the Supp. Material, which includes Refs. \cite{Banszerus2020Dec, Knothe2020Jun, lemonik_competing_2012, lemonik_spontaneous_2010, ochoaSpinorbitCouplingAssisted2012, xiaoBerryPhaseEffects2010, moulsdaleEngineeringTopologicalMagnetic2020, parkValleyFilteringDue2017, Fuchs2010}.}. We obtain the tunnel rates for a single-electron sequential tunnelling process to first order in the tunnelling Hamiltonian and by applying Fermi's golden rule. Using the tunnel rates we compute the occupation probabilities for the different QD states in the stationary limit and the resulting sequential tunnelling current. Employing this procedure and adjusting the free parameters $g^{(1)}_v$, $g^{s,a}_v$,  $\delta_{1,2}$, $t^{S,D}$ and $\Delta_\mathrm{SO}$, we obtain differential transconductance maps~\cite{Knothe2021Apr} as in Fig.~2(b).

Now, we assign the most prominent features in Figs.~2(a) and (b) to their corresponding transitions from single- to two-particle states. Fig.~2(c) shows all possible transitions at zero magnetic field, where the spin-orbit splitting was neglected for simplicity.
The length of each transition arrow directly corresponds to the chemical potential required for that transition. 
Identifying transitions with the features in Fig.~2(a) allows to extract the involved energy scales: 
The orbital splitting is $\Delta_\mathrm{Orb} \approx 700\,\mu$eV, with the symmetric orbital providing the GS transition (5,6) and the anti-symmetric orbital appearing at position (I) with transitions 1,3,8 and 10. From the inset in Fig.~2(a), we can extract $\delta_2 \approx 350\,\mu$eV, while $\delta_1 \approx 0$ within the measurement resolution.
Fig.~2(d) shows the level scheme at finite $B_\perp$. We can identify three different slopes that match neatly with the data. The first ones are caused by transitions where only the single-particle states shift with $g_v^{(1)}$ and the two-particle states remain constant (3-8), while the other two are due to transitions from the single-particle states to the valley polarized triplets in either the symmetric (2,9) or anti-symmetric (1,10) orbital \footnote{$\ket{K^+}$ is an ES of the QD at $\mathcal{N} = 1$ and finite $B_\perp$, which makes its occupations probability smaller than $\ket{K^-}$. Thus, most transitions originating from $\ket{K^+}$ are not clearly visible in the data, as their contribution to the transport current is much smaller.}. The change in chemical potential for each transition is given by 
$\Delta \mu_{3-8} = \pm \frac{1}{2} g^{(1)}_v \mu_B B_\perp$, $\Delta \mu_{2,9} = \pm  (g^s_v -  \frac{1}{2} g^{(1)}_v) \mu_B B_\perp\,$ and
$\Delta \mu_{1,10} = \pm (g^a_v - \frac{1}{2} g^{(1)}_v ) \mu_B B_\perp$, respectively.

Taking these different slopes into account, it can be understood how the GS transition evolves with magnetic field. Transitions 5 and 6 are the GS transition for zero magnetic field. For increasing positive magnetic field, transition 6 requires more chemical potential, while transition 5 needs less. Still, transition 6 remains the GS transition (which defines the Coulomb blockaded region), since $\ket{\text{K}^-}$ is the single-particle GS. Instead, transition 5 becomes a 'negative' excited state, which manifests as a decrease in transconductance (see arrow 5 in Fig.~2(a)). At position (II), transition 9 becomes the GS transition, when $\ket{S^s \, T^v_-}$ becomes the two-particle GS. This reveals that the symmetric orbital is lower in energy than the anti-symmetric one, since only the multiplet splitting of the symmetric orbital gives rise to this change in the GS transition. 
When further increasing the magnetic field, transition 10 eventually becomes the GS transition (see position (III)), as soon as $\ket{T^s_{0\pm} \, T^v_-}$ is lower in energy than $\ket{S^s \, T^v_-}$, showing that $g_v^a > g_v^s$.

In order to better understand the energy scales involved in the two-particle spectrum, the measurement of Fig.~2(a) is performed for different displacement fields, $D$, applied to the BLG, which changes the band structure and also the confinement potential of the QD \cite{Knothe2020Jun}. For each displacement field, we evaluated $\Delta_\mathrm{Orb}, \delta_2$, and the valley $g$-factors of the most distinct transitions.  For transitions 3 -- 8 in Fig.~2, the slope only arises from the single-particle states, since the two-particle states do not shift with magnetic field. Fig.~3(a) shows that the single-particle $g$-factor evaluated from transitions 6 and 8 are the same within the error margins, even though they target states in two different orbitals. There is a decrease of $g^{(1)}_v$ for higher displacement fields, which was also observed earlier~\cite{Tong2021Jan}. 
This decrease is also visible in Fig.~3(b), where the $g$-factors of the symmetric and anti-symmetric orbitals $g^{s,a}_v$, are evaluated from transitions 9 and 10. They show an average difference of  $g^{a}_v - g^{s}_v \approx 8$, confirming that the symmetric and anti-symmetric orbital comprise of different single-particle orbital states.
The decrease of all $g$-factors in Fig.~3 indicates a shift in the wavefunctions' composition in $k$-space. This conclusion is supported by the fact that both the orbital splitting as well as the short range interaction contribution $\delta_2$ show the same trend in Fig.~3(c), since they also scale with the shape of the wavefunction \cite{Knothe2020Jun}. The orbital splitting,  $\Delta_\mathrm{Orb}$,  decreases from $\approx 0.8$ to $0.4$~meV for increasing displacement fields, while $\delta_2$ decreases from $\approx 0.5$ to $0.15$~meV. For all displacement fields, $\delta_1$ $\approx 0$ within the measurement resolution. 

The measurements of the splittings in Fig.~3(c) allow conclusions about the microscopic short-range interaction constants in BLG.
We can relate the multiplet splittings $\delta_{1,2}$  to the short-range coupling constants~\cite{lemonik_competing_2012, lemonik_spontaneous_2010, Knothe2021Apr}, $g_\perp$, (quantifying inter-valley scattering) and, $g_{z0}, g_{0z}$, (generated by "current-current" interactions): $\delta_1 + \delta_2 = 8|g_\perp\mathfrak{J}|$ and $\delta_2 - \delta_1 = 4|(g_{z0} + g_{0z})\mathfrak{J}|$.
\rev{The influence of the QD size, confinement potential and band gap on the short-range splitting is captured in $\mathfrak{J}$, which is the wavefunction overlap of the specific QD state~\cite{Knothe2021Apr, Knothe2020Jun}}.    
Consequently, our experimental observation of near-vanishing $\delta_1$ indicates that inter-valley scattering and "current-current" interactions are of the same magnitude in BLG QDs, i.e., $2|g_{\perp}| \approx |g_{z0} + g_{0z}|$. 
Calculating $\mathfrak{J}$ for QDs of radius $\approx 80$~nm yields $\mathfrak{J}\approx 4\cdot10^{−4}$ nm$^{-2}$ \cite{Knothe2020Jun}. 
Combining this with our experimental results for $\delta_2$,  we obtain an estimate for the short-range BLG coupling constants~\footnote{We cannot determine the sign of the coupling constants from comparison to the transport calculations. 
Changing the sign of $g_{\perp}, g_{z0}, g_{0z}$ reverses the order of states in the multiplets shown in Figs.~2(c,d) but leaves the pattern intact and hence leads to the same differential conductance maps as in Figs.~2(b).}, $|g_{\perp}|\approx 0.08$~eVnm$^2$, which is in accordance with the order of magnitude estimated previously from microscopic calculations \cite{Knothe2020Jun}.


%

In summary, we have experimentally observed both orbitally symmetric and anti-symmetric two-particle states in a BLG QD using finite bias tunneling spectroscopy. We identified that the 16 possible states are split into orbitally symmetric and anti-symmetric two-particle states separated by $\Delta_\mathrm{Orb}\approx 0.4 - 0.8$~meV. The orbitally symmetric multiplet is further split with $\delta_2\approx 0.15 - 0.5$~meV by lattice scale interactions which are equally related to inter-valley scattering and "current-current" interactions. \rev{We find that the two-particle ground state is a spin-triplet at $B=0$ but can be tuned to be a spin-singlet for finite out-of-plane magnetic field. This is in contrast to semiconductor QDs, where the spin-singlet is the two-particle ground state for common magnetic field strengths and is utilized for qubit read-out in double QDs via Pauli spin blockade. Understanding the two-particle spectrum in BLG QDs is thus an essential step for further investigating the influence of the valley on Pauli blockade in double QDs and eventually for identifying a suitable regime for qubit operations.}\\



\textbf{Data availability}
The data supporting the findings of this study are available in a Zenodo repository under \url{https://doi.org/10.5281/zenodo.5788690}.

\textbf{Acknowledgments} 
This project has received funding from the European Union's Horizon 2020 research and innovation programme under grant agreement No. 881603 (Graphene Flagship) and from the European Research Council (ERC) under grant agreement No. 820254, the Deutsche Forschungsgemeinschaft (DFG, German Research Foundation) under Germany's Excellence Strategy - Cluster of Excellence Matter and Light for Quantum Computing (ML4Q) EXC 2004/1 - 390534769, through DFG (STA 1146/11-1), and by the Helmholtz Nano Facility~\cite{Albrecht2017May}. VIF and AK were supported by EC-FET Core 3 European Graphene Flagship Project, EC-FET Quantum Flagship Project 2D-SIPC, and Lloyd Register Foundation Nanotechnology Grant. LIG was supported by the NSF DMR-2002275. Growth of hexagonal boron nitride crystals was supported by the Elemental Strategy Initiative conducted by the MEXT, Japan, Grant Number JPMXP0112101001,  JSPS KAKENHI Grant Numbers JP20H00354 and the CREST(JPMJCR15F3), JST.

\bibliography{Literature}


\end{document}


\author{S.~M\"oller}
\thanks{These two authors contributed equally.}
\author{L.~Banszerus}
\thanks{These two authors contributed equally.}
\affiliation{JARA-FIT and 2nd Institute of Physics, RWTH Aachen University, 52074 Aachen, Germany,~EU}%
\affiliation{Peter Gr\"unberg Institute  (PGI-9), Forschungszentrum J\"ulich, 52425 J\"ulich,~Germany,~EU}
\author{A. Knothe}
\affiliation{National Graphene Institute, University of Manchester, Manchester M13 9PL, United Kingdom}%
\author{C.~Steiner}
\affiliation{JARA-FIT and 2nd Institute of Physics, RWTH Aachen University, 52074 Aachen, Germany,~EU}
\author{E.~Icking}
\affiliation{JARA-FIT and 2nd Institute of Physics, RWTH Aachen University, 52074 Aachen, Germany,~EU}%
\affiliation{Peter Gr\"unberg Institute  (PGI-9), Forschungszentrum J\"ulich, 52425 J\"ulich,~Germany,~EU}
\author{S.~Trellenkamp}
\author{F.~Lentz}
\affiliation{Helmholtz Nano Facility, Forschungszentrum J\"ulich, 52425 J\"ulich,~Germany,~EU}

\author{K.~Watanabe}
\affiliation{Research Center for Functional Materials, 
National Institute for Materials Science, 1-1 Namiki, Tsukuba 305-0044, Japan}
\author{T.~Taniguchi}
\affiliation{International Center for Materials Nanoarchitectonics, 
National Institute for Materials Science,  1-1 Namiki, Tsukuba 305-0044, Japan}%
\author{L.~I.~Glazman}
\affiliation{Departments of Physics and Applied Physics, Yale University, New Haven, CT 06520, USA}
\author{V. Fal'ko}
\affiliation{National Graphene Institute, University of Manchester, Manchester M13 9PL, United Kingdom}
\author{C.~Volk}
\author{C.~Stampfer}
\email{stampfer@physik.rwth-aachen.de}
\affiliation{JARA-FIT and 2nd Institute of Physics, RWTH Aachen University, 52074 Aachen, Germany,~EU}%
\affiliation{Peter Gr\"unberg Institute  (PGI-9), Forschungszentrum J\"ulich, 52425 J\"ulich,~Germany,~EU}%

\title{Supporting information: Probing two-electron multiplets in bilayer graphene quantum dots}

\date{\today}
\maketitle

\section{Details on Calculations}
\label{sec:CalcDetails}
We outline the structure of the \BLG{} quantum dot's single and two-particle states that we have identified in the experiment. \\
 We characterise a one-electron state in the $\SPorbital$-th dot orbital with the spin ($\SPspin=\uparrow, \downarrow$) and valley ($\SPvalley=+,-$) degree of freedom and write $|\SPorbital,\SPspin,\SPvalley\rangle=\dotc_{\SPorbital \SPspin \SPvalley}|0\rangle$, where $\dotc_{\SPorbital \SPspin \SPvalley}$ is the electron creation operator and $|0\rangle$ is the empty dot state. The \SP{} states' energy in a perpendicular magnetic field, $B_{\perp}$, is
\begin{equation}
E_{\SPorbital,\SPspin=\uparrow, \downarrow,\SPvalley}=E_{\SPorbital}+ \Ecap(1)\pm \frac{1}{2}\SPvalley\Delta_\mathrm{SO}\pm \frac{1}{2} \gS\mu_{B} B_{\perp} +\frac{1}{2} \SPvalley \gVSP\mu_{B} B_{\perp}.
\label{eqn:SPEn}
\end{equation}
Here,  $E_{\SPorbital}$ is the energy of the $\SPorbital$-th orbital dot state at zero magnetic field, $\Delta_\mathrm{SO}$ is the Kane-Mele spin-orbit splitting (enhanced by zero-point vibrations [39]),  $\gS=2$ is the free electron spin $g$-factor,  $\gVSP$ is the valley $g$-factor (a consequence of gapped \BLG{}'s nontrivial Bloch band Berry curvature entailing an topological orbital magnetic moment  with opposite sign in the two different valleys~[40-43], and $\mu_{B}$ is the Bohr magneton. We account for  the finger gate with capacitance $\CapC$ in the experimental setup, which, at gate voltage $\Vgate$, changes the dot's electrostatic potential by,
\begin{equation}
\Ecap(\partnumb)=\frac{(\partnumb e - \CapC\;  \Vgate)^{2}}{2e C},
\label{eqn:Ecap}
\end{equation}
with $\partnumb$ being the dot occupation number and $C$  the total capacitance of the dot.\\
In the experiment, we identify the lowest energy \TP{} state to be orbitally symmetric ($\phi_s$), while the first \TP{} state features an antisymmetric ($\phi_a$) orbital wave function.
There are six combinations of spin ($s$) and valley ($v$) -singlet (S)   and -triplet (T)  states and an orbitally symmetric \TP{} state of orbital $\SPorbital$:
\begin{align}
&\nonumber\StatenSzeroTplusX=\frac{1}{\sqrt{2}}(\dotc_{\SPorbital\upplus}\dotc_{\SPorbital\downminus}-\dotc_{\SPorbital\downplus}\dotc_{\SPorbital\upminus})|0\rangle,\\
&\nonumber\StatenSzeroTminusZ=\dotc_{\SPorbital\upminus}\dotc_{\SPorbital\downminus}|0\rangle,\\
&\nonumber\StatenSzeroTplusZ=\dotc_{\SPorbital\upplus}\dotc_{\SPorbital\downplus}|0\rangle,\\
&\nonumber\StatenmminusOneTminusX=\dotc_{\SPorbital\downplus}\dotc_{\SPorbital\downminus}|0\rangle,\\
&\nonumber\StatenmZeroTminusX=\frac{1}{\sqrt{2}}(\dotc_{\SPorbital\upplus}\dotc_{\SPorbital\downminus}+\dotc_{\SPorbital\downplus}\dotc_{\SPorbital\upminus})|0\rangle,\\
&\StatenmplusOneTminusX=\dotc_{\SPorbital\upplus}\dotc_{\SPorbital\upminus}|0\rangle.
\label{eqn:sStates}
\end{align}
The energies of this \TP{} \GS{} multiplet in a finite perpendicular magnetic field $B_{\perp}$ are [17], 
\begin{align}
\nonumber &E_{\LnnSzeroTplusX}=\TPEn^{\TPsymm}_{\SPorbital\SPorbital} +(g_{zz}+ 4g_{\perp}-g_{0z}-g_{z0})\intW  + \Ecap(2) ,\\
\nonumber &E_{\LnnSzeroTminusZ} =\TPEn^{\TPsymm}_{ \SPorbital\SPorbital} + (g_{zz} +g_{0z}+g_{z0})\intW + \Ecap(2) -\gVTPs \mu_{B} B_{\perp},\\
\nonumber &E_{\LnnSzeroTplusZ} =\TPEn^{\TPsymm}_{ \SPorbital\SPorbital} + (g_{zz} +g_{0z}+g_{z0})\intW + \Ecap(2) +\gVTPs\mu_{B} B_{\perp},\\
\nonumber &E_{\LnnmminusOneTminusX} =\TPEn^{\TPsymm}_{ \SPorbital\SPorbital} + (g_{zz}-4g_{\perp}-g_{0z}-g_{z0})\intW + \Ecap(2) -\gS\mu_{B} B_{\perp},\\
\nonumber &E_{\LnnmZeroTminusX} =\TPEn^{\TPsymm}_{ \SPorbital\SPorbital} + (g_{zz}-4g_{\perp}-g_{0z}-g_{z0})\intW + \Ecap(2) ,\\
& E_{\LnnmplusOneTminusX} =\TPEn^{\TPsymm}_{ \SPorbital\SPorbital} + (g_{zz}-4g_{\perp}-g_{0z}-g_{z0})\intW + \Ecap(2) +\gS \mu_{B}B_{\perp}.
 \label{eqn:sEnergies}
\end{align}
 Here,  $\TPEn^{\TPsymm}_{ \SPorbital\SPorbital}$, comprises the energy of the $\SPorbital$-th \SP{} orbital and the screened long-range electron-electron Coulomb interaction. Since any symmetric orbital wave function has finite density at zero or small inter-particle real space distances, these states are further affected by short-range (lattice-constant-scale) electron-electron interactions [19, 20]. 
 The factor, 
 \begin{equation}
     \intW=\int d\mathbf{r}[\psi_{\SPorbital\SPspin_1\SPvalley_1}^{B^{\prime}}(\mathbf{r})]^*[\psi_{\SPorbital\SPspin_2\SPvalley_2}^{B^{\prime}}(\mathbf{r})][\psi_{\SPorbital\SPspin_3\SPvalley_3}^{B^{\prime}}(\mathbf{r})]^*[\psi_{\SPorbital\SPspin_4\SPvalley_4}^{B^{\prime}}(\mathbf{r})]>0
 \end{equation}
 (for all combinations of $\SPvalley_i$ corresponding to  inter- and intra-valley scattering channels), is computed from the wave functions, $\psi_{\SPorbital\SPspin\SPvalley}^{B^{\prime}}$, on the occupied \BLG{} sublattice $B^{\prime}$ (a finite diplacement field forces layer polarization), and hence   depends on the experimental dot state characteristics, i.e.,  dot shape, gap, and mode number.  
 The valley g-factor, $\gVTPs$, of the \TP{} states is given by the sum of the \SP{} g-factors in the two valleys. 
 
For the first excited \TP{} dot state that we observe in experiment, the ten possible \TP{} states with orbitally antisymmetric (\TPasymm)  wave function are,
\begin{align}
\nonumber \StatenmmminusOneTminusZ=&\dotc_{\SPorbital\downminus}\dotc_{\SPorbitalm\downminus}|0\rangle,\\
\nonumber \StatenmmZeroTminusZ=&\frac{1}{\sqrt{2}}(\dotc_{\SPorbital\upminus}\dotc_{\SPorbitalm\downminus}+\dotc_{\SPorbital\downminus}\dotc_{\SPorbitalm\upminus})|0\rangle,\\
\nonumber \StatenmmplusOneTminusZ=&\dotc_{\SPorbital\upminus}\dotc_{\SPorbitalm\upminus}|0\rangle,\\
\nonumber  \StatenmmminusOneTplusX=&\frac{1}{\sqrt{2}}(\dotc_{\SPorbital\downplus}\dotc_{\SPorbitalm\downminus}+\dotc_{\SPorbital\downminus}\dotc_{\SPorbitalm\downplus})|0\rangle,\\
\nonumber \StatenmmZeroTplusX=&\frac{1}{2}(\dotc_{\SPorbital\upplus}\dotc_{\SPorbitalm\downminus}+\dotc_{\SPorbital\upminus}\dotc_{\SPorbitalm\downplus}+\dotc_{\SPorbital\downplus}\dotc_{\SPorbitalm\upminus}+\dotc_{\SPorbital\downminus}\dotc_{\SPorbitalm\upplus})|0\rangle,\\
\nonumber \StatenmmplusOneTplusX=&\frac{1}{\sqrt{2}}(\dotc_{\SPorbital\upplus}\dotc_{\SPorbitalm\upminus}+\dotc_{\SPorbital\upminus}\dotc_{\SPorbitalm\upplus})|0\rangle,\\
\nonumber \StatenmmminusOneTplusZ=&\dotc_{\SPorbital\downplus}\dotc_{\SPorbitalm\downplus}|0\rangle,\\
\nonumber \StatenmmZeroTplusZ=&\frac{1}{\sqrt{2}}(\dotc_{\SPorbital\upplus}\dotc_{\SPorbitalm\downplus}+\dotc_{\SPorbital\downplus}\dotc_{\SPorbitalm\upplus})|0\rangle,\\
\nonumber \StatenmmplusOneTplusZ=&\dotc_{\SPorbital\upplus}\dotc_{\SPorbitalm\upplus}|0\rangle,\\
 \StatenmSzeroTminusX =&\frac{1}{2}(\dotc_{\SPorbital\upplus}\dotc_{\SPorbitalm\downminus}-\dotc_{\SPorbital\upminus}\dotc_{\SPorbitalm\downplus}-\dotc_{\SPorbital\downplus}\dotc_{\SPorbitalm\upminus}+\dotc_{\SPorbital\downminus}\dotc_{\SPorbitalm\upplus})|0\rangle.
\label{eqn:asStates}
\end{align}
Hence the excited \TP{} state energies are given by,
\begin{align}
\nonumber& E_{\LnmmminusOneTminusZ}=\TPEn^{\TPasymm}_{ \SPorbital \SPorbitalm} -\gVTPa \mu_{B} B_{\perp} + \Ecap(2) +\Delta_{SO}-\gS \mu_{B} B_{\perp} ,\\
\nonumber& E_{\LnmmZeroTminusZ}=\TPEn^{\TPasymm}_{ \SPorbital\SPorbitalm} + \Ecap(2) -\gVTPa \mu_{B}B_{\perp} ,\\
\nonumber& E_{\LnmmplusOneTminusZ}  =\TPEn^{\TPasymm}_{ \SPorbital\SPorbitalm}-\gVTPa \mu_{B}B_{\perp} + \Ecap(2) -\Delta_{SO}+\gS\mu_{B} B_{\perp} ,\\
\nonumber& E_{\LnmmminusOneTplusX}=\TPEn^{\TPasymm}_{ \SPorbital\SPorbitalm} + \Ecap(2) -\gS \mu_{B} B_{\perp}  ,\\
\nonumber& E_{\LnmmZeroTplusX}=\TPEn^{\TPasymm}_{ \SPorbital\SPorbitalm} + \Ecap(2)  ,\\
\nonumber &E_{\LnmmplusOneTplusX}=\TPEn^{\TPasymm}_{\SPorbital\SPorbitalm}  + \Ecap(2) +\gS \mu_{B} B_{\perp} ,\\
\nonumber &E_{\LnmmminusOneTplusZ} =\TPEn^{\TPasymm}_{ \SPorbital\SPorbitalm} +\gVTPa \mu_{B}B_{\perp} + \Ecap(2) -\Delta_{SO}-\gS \mu_{B} B_{\perp} ,\\
\nonumber &E_{\LnmmZeroTplusZ}=\TPEn^{\TPasymm}_{ \SPorbital\SPorbitalm} +\gVTPa \mu_{B}B_{\perp} + \Ecap(2)  ,\\
\nonumber &E_{\LnmmplusOneTplusZ}  =\TPEn^{\TPasymm}_{ \SPorbital\SPorbitalm}  +\gVTPa \mu_{B} B_{\perp} + \Ecap(2)+\Delta_{SO} +\gS \mu_{B} B_{\perp},\\
& E_{\LnmSzeroTminusX}=\mathfrak{E}^{\TPasymm}_{\SPorbital\SPorbitalm}  + \Ecap(2) .
\label{eqn:asEnergies}
\end{align}
Here, $\TPEn^{\TPasymm}_{ \SPorbital\SPorbitalm}$ is the energy of the orbitally antisymmetric states of two screened interacting electrons in \SP{} orbitals $\SPorbital$ and $\SPorbitalm$ (akin to the orbitally symmetric state described above),  
and   $\gVTPa$  is the  valley g-factor of the \TP{} multiplet.
\\

We model the leads in the experiment as single channels and a corresponding lead-dot tunnelling Hamiltonian,
\begin{align}
\HLeads&=\sum_{l=L,R}\sum_{k,\SPspin,\SPvalley}\epsilon^{l}_k\leadc_{lk\SPspin\SPvalley}\leada_{lk\SPspin\SPvalley},\\
\HTun&=\sum_{l=L,R}\sum_{n,k,\SPspin,\SPvalley} \big(t^{l}_{\SPorbital\SPspin\SPvalley}\leadc_{lk\SPspin\SPvalley}\dota_{\SPorbital\SPspin\SPvalley} +\text{h.c.}\big),
\end{align}
where $\leadc_{lk\SPspin\SPvalley}$ creates a lead electron with momentum k, energy $\epsilon^{l}_{k}$, spin $\SPspin$, and valley quantum number  $\SPvalley$), and $t^L$ ($t^R$) is the tunnelling amplitude for the left (right) lead.
The experimentally observed transport features can be described by single-electron sequential tunnelling. We obtain the corresponding tunnelling rates by expanding to first order in the tunnelling Hamiltonian:
\begin{widetext}
\begin{equation}
W_{2:\chi\leftarrow1:\chi^{\prime}}=\sum_{{l=L,R}}W^{l}_{2:\chi\leftarrow1:\chi^{\prime}}=\sum_{{l=L,R}}\frac{2\pi}{\hbar}|t^{l}_{\SPorbital\SPspin\SPvalley}|^{2}f(E_{2:\chi}-E_{1:\chi^{\prime}}-\Chempot^{l}),
\label{eqn:WSeqT}
\end{equation}
\end{widetext}
With $f$ being the Fermi function and $\Chempot^{l}$ the electro-chemical potential of lead $l$. We consider the case of symmetric bias, $\Chempot^{L/R}=\pm e \Vbias$, with respect to the equilibrium electro-chemical potential. Depending on the dot occupation number ($\partnumb=1$ or $\partnumb=2$) $\chi$ indexes the states from the \SP{} or \TP{} multiplets, respectively.\\
We solve the master equation of the probabilities, $\dotprob_{\dotstate}$, for the state, $|\dotstate\rangle$, to be occupied at a given time,
\begin{equation}
\dot{\dotprob}_{\dotstate}=\sum_{\dotstateP}(W_{\dotstate\leftarrow \dotstateP}\dotprob_{\dotstateP}-W_{\dotstateP\leftarrow \dotstate}\dotprob_{\dotstate}),
\label{eqn:rateeqn}
\end{equation}
in the stationary limit, $\dot\dotprob_{\dotstate}=0$, using the normalization condition $\sum_{\dotstate}\dotprob_{\dotstate}=1$.
From the probabilities we compute the total particle current flowing from the dot to lead $l$,
\begin{equation} 
 I^{l}=\sum_{1:\chi,2:\chi^{\prime}} (W^l_{1:\chi \leftarrow 2:\chi^{\prime}})e\dotprob_{2:\chi^{\prime}}-(W^l_{2:\chi^{\prime}\leftarrow1:\chi})e\dotprob_{1:\chi},
 \label{eqn:Iseq}
\end{equation}
and the differential conductance, $dI/dV_G$.

\newpage
\section{Additional simulated data}
\begin{figure}[!thb]
\includegraphics[draft=false,keepaspectratio=true,clip,width=\linewidth]{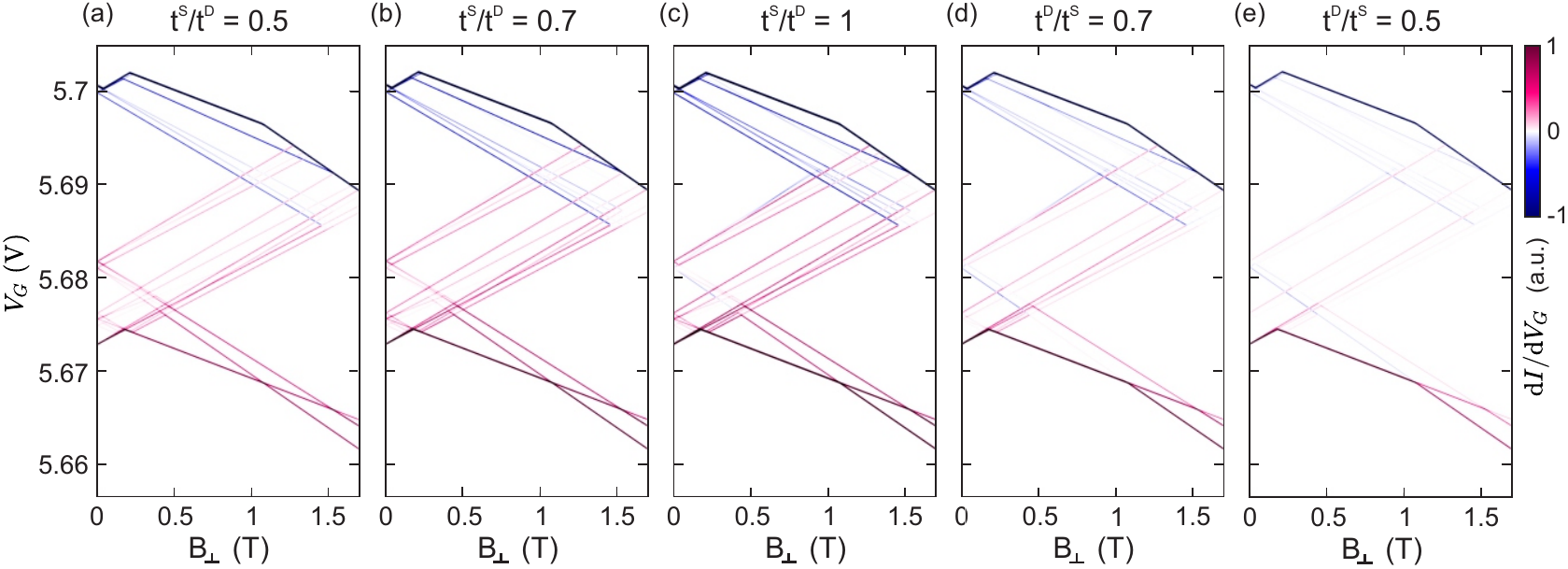}
\caption[Fig01]{\textbf{(a - e)} Theoretically calculated differential transconductance $dI/dV_{G}$ as a function of $V_G$ and perpendicular magnetic field, as in Fig.~2(b). Different tunnel couplings to source and drain change the prominence of line features.}
\label{fig:SuppDataTheo}
\end{figure}

We compute the charge current across the dot, \autoref{eqn:Iseq}, and the differential conductance for various system parameters of the dot and the leads. We take values for the spin- and valley g-factors, the orbital energies, and the short-range interaction coupling parameters from the experiment, as explained in the main text. \autoref{fig:SuppDataTheo} shows calculated differential conductance maps for different values of tunnelling amplitudes to the left (source contact in our convention) and right (drain) contacts. 

\newpage
\section{Additional experimental data}
\begin{figure}[!thb]
\includegraphics[draft=false,keepaspectratio=true,clip,width=\linewidth]{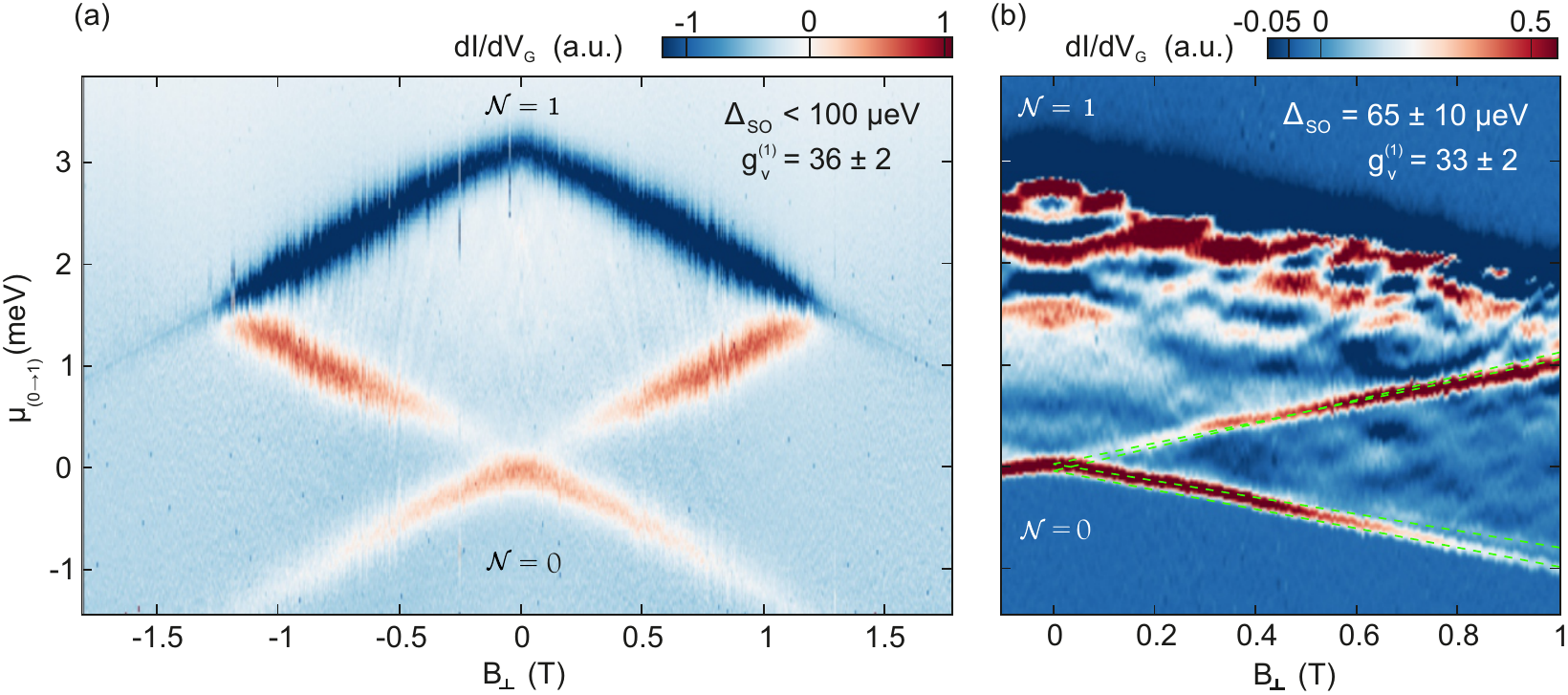}
\caption[Fig01]{\textbf{(a, b)} Differential transconductance $dI/dV_{G}$ as a function of $V_G$ (transformed to electro-chemical potential) and perpendicular magnetic field, measured across the $\mathcal{N} = 1$ Coulomb peak at $V_{b} = 3$~mV similar to Fig.~2(a). For the first electron, the electro-chemical potential of the transition directly corresponds to the energy of the single particle state occupied, which confirms that the single particle spectrum presented in Fig.~1(c) is indeed observed. (a) shows data from the device presented in the main manuscript, (b) shows data from a different device where the spin-orbit splitting could be resolved.}
\label{f1}
\end{figure}

 \begin{figure}[!thb]
\includegraphics[draft=false,keepaspectratio=true,clip,width=0.8\linewidth]{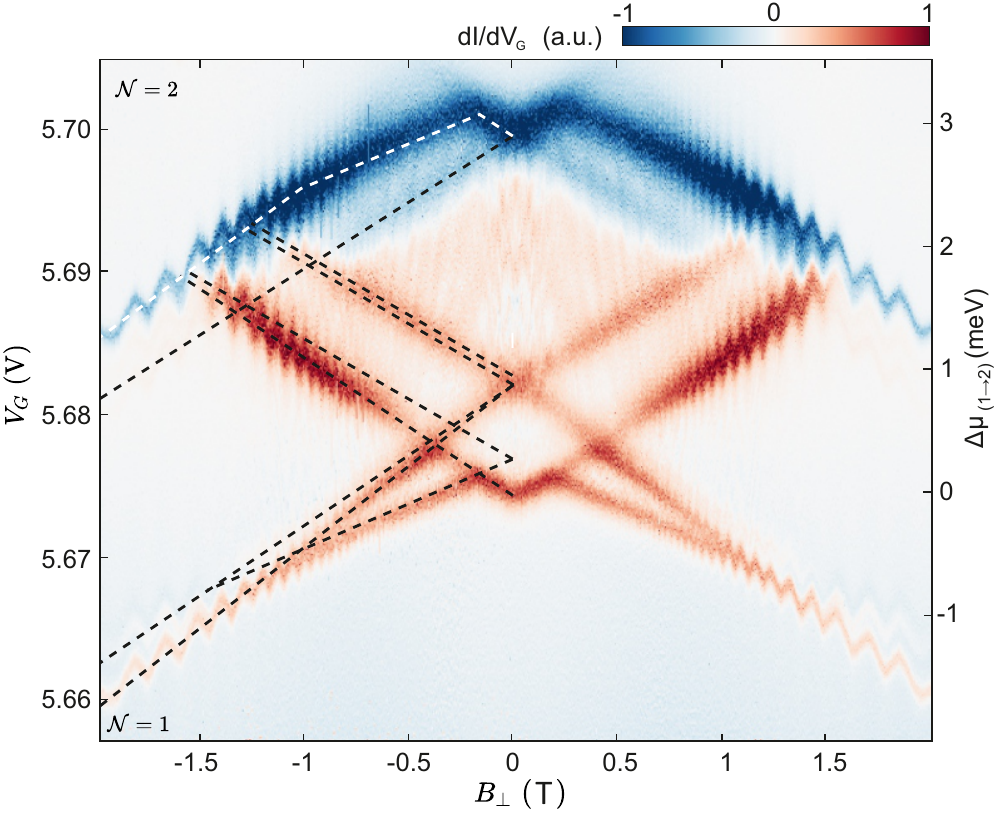}
\caption[Fig01]{
\textbf{(a)} Raw data from Fig.~2(a). Differential transconductance $dI/dV_{G}$ as a function of FG voltage and perpendicular magnetic field, measured across the $\mathcal{N} = 2$ Coulomb peak at $V_{b} = 3$~mV. Shubnikov-de-Haas oscillations of the density of states in source and drain lead to periodic changes of the lever-arm, which oscillates with $1/B$ [25]. }
\label{f1}
\end{figure}

\begin{figure}[!thb]
\includegraphics[draft=false,keepaspectratio=true,clip,width=\linewidth]{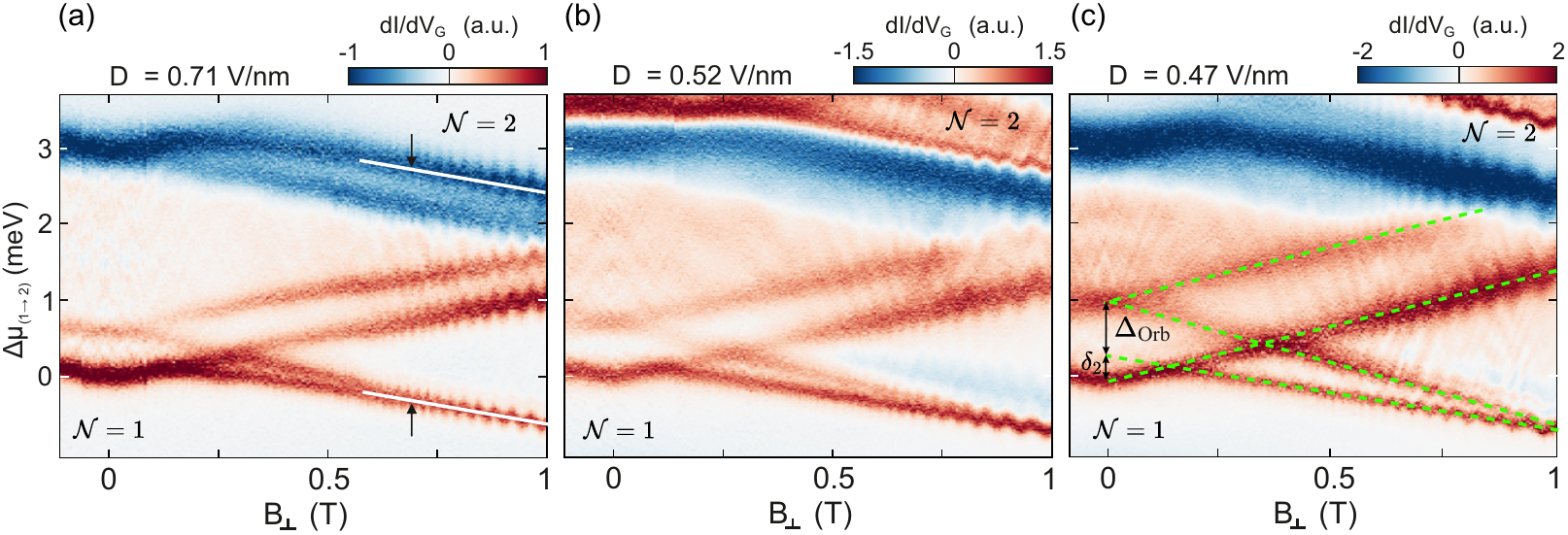}
\caption[Fig01]{\textbf{(a - c)} Differential transconductance $dI/dV_{G}$ as a function of $V_G$ (transformed to electro-chemical potential) and perpendicular magnetic field, measured across the $\mathcal{N} = 1$ Coulomb peak at $V_{b} = 3$~mV similar to Fig.~2(a) but for different displacement fields. The white lines in (a) correspond to the applied bias via $e V_b = \Delta \mu$, which allows to compute the lever-arm of the FG.
The dashed lines and arrows in (c) illustrate the evaluation of the energy scales and $g$-factors. The slight trend of larger $\Delta_\mathrm{Orb}$ and $\delta_2$ for smaller displacement fields is nicely visible.}
\label{f1}
\end{figure}

\rev{In this section we provide additional experimental data.
In Fig.~S2 we show differential transconductance ($dI/dV_{G}$) measurements as a function of the chemical potential (obtained from $V_\mathrm{G}$ and the experimentally extracted gate lever-arm) and $B_\perp$ for two different BLG QD devices. The data are recorded across the $\mathcal{N} = 1$ Coulomb peak at $V_{b} = 3$~mV (see also Fig.~1(b) in the main manuscript).
In Fig.~S3 we show the measured raw data of Fig.~2(a) in the main manuscript, where the Shubnikov-de-Haas oscillations at high $B_\perp$ are well visible. 
Fig.~S4 illustrates how the data points in Fig.~3 in the main manuscript are obtained. First, we fix a voltage for the back gate (BG) and measure the conductance as a function of the voltages applied to the two split gates (SGs), in order (i) to properly pinch-off the BLG underneath the split gates and (ii) to adjust the displacement field $D$. BG and SG voltages for the pinch-off do not scale perfectly linear, which makes this procedure necessary for every single data point. Then, we measure across the second Coulomb peak ($\mathcal{N} = 2$) at finite bias ($V_b = 3 \,$mV) as a function of out-of-plane magnetic field to obtain data sets like Fig.~S4(a)-(c). Next, we determine the lever-arm by the outline of the conducting region, which corresponds to $\Delta \mu = e V_b$ (see white lines in Fig.~S4(a)). Finally, we obtain the various $g$-factors from the different slopes (see dashed lines in Fig.~S4(c)) and we extract $\Delta_{\text{Orb}}$ as well as $\delta_2$ as shown by the black arrows in Fig.~S4(c).}